\documentclass[preprint2]{aastex}

\usepackage{epsf}
\usepackage{graphics}

\newcommand{\kms}{km s$^{-1}$}

\newcommand{\ax}{$\alpha_{\rm X}$}

\newcommand{\rb}[1]{\raisebox{1.5ex}[-1.5ex]{#1}}
\newcommand{\msun}{$M_{\odot}$}
\newcommand{\dM}{$\dot M$}

\newcommand{\mbh}{$M_{\rm BH}$}
\newcommand{\rs}{$r_{\rm S}$}
\newcommand{\ts}{$t_{\rm S}$}
\newcommand{\lx}{$L_{\rm X}$}
\shorttitle{Soft X-ray AGN: II. Statistics}
\shortauthors{Grupe}

\begin{document}

\def\clipfig#1{\def\lbracket{[}\def\testit{#1}%
    \ifx\testit\lbracket\let\next=\optclipfig\else\let\next=\stdclipfig\fi%
    \next{#1}}
%
\newcommand {\hclipfig} [7] {\clipfig[#7]{#1}{#2}{#3}{#4}{#5}{#6}}
%
\def\usemodepsfig {\global\def\cfmode{x}\typeout{*** set clipfig to PSFIG mode ***}}
\def\usemodeepsf  {\global\def\cfmode{}\typeout{*** set clipfig to EPSF mode ***}}
\def\useunitmm    {\global\def\cfunit{x}\typeout{*** set clipfig to use mm as unit ***}}
\def\useunitcm    {\global\def\cfunit{}\typeout{*** set clipfig to use cm as unit ***}}
\def\clipfigsettings {\ifx\cfmode\empty\def\ccfmode{EPSF }\else\def\ccfmode{PSFIG }\fi%
    \ifx\cfunit\empty\def\ccfunit{cm }\else\def\ccfunit{mm }\fi%
    \typeout{*** current clipfig settings: \ccfmode mode, using \ccfunit as unit ***}}
%
%
%
%
\def\stdclipfig#1#2#3#4#5#6{\ifx\cfmode\empty%
    \let\next=\eclipfig\else\let\next=\pclipfig\fi%
    \next{#1}{#2}{#3}{#4}{#5}{#6}}
\def\optclipfig#1#2]#3#4#5#6#7#8{\ifx\cfmode\empty%
    \let\next=\ehclipfig\else\let\next=\phclipfig\fi%
    \next{#3}{#4}{#5}{#6}{#7}{#8}{#2}}
%
%
%
\newcommand {\pclipfig}[6] {\ifx\cfunit\empty%
        \psfig{figure=#1.ps,width=#2cm,bbllx=#3cm,bblly=#4cm,bburx=#5cm,%
           bbury=#6cm,clip=}\else%
        \psfig{figure=#1.ps,width=#2mm,bbllx=#3mm,bblly=#4mm,bburx=#5mm,%
           bbury=#6mm,clip=}\fi}
\newcommand {\phclipfig}[7] {\ifx\cfunit\empty%
        \hspace{#7cm}\psfig{figure=#1.ps,width=#2cm,bbllx=#3cm,bblly=#4cm,%
           bburx=#5cm,bbury=#6cm,clip=}\else%
        \hspace{#7mm}\psfig{figure=#1.ps,width=#2mm,bbllx=#3mm,bblly=#4mm,%
           bburx=#5mm,bbury=#6mm,clip=}\fi}
%
%
%
\newcommand {\eclipfig}[6]{%
  \ifx\cfunit\empty\epsfxsize=#2cm\else\epsfxsize=#2mm\fi%
  \epsfclipon\epsfverbosetrue%
  \cfcmtopspts{#3}\cfllxi=\cftempi\cfllxf=\cftempf%
  \cfcmtopspts{#4}\cfllyi=\cftempi\cfllyf=\cftempf%
  \cfcmtopspts{#5}\cfurxi=\cftempi\cfurxf=\cftempf%
  \cfcmtopspts{#6}\cfuryi=\cftempi\cfuryf=\cftempf%
  \def\cfstra{\number\cfllxi.\number\cfllxf}%
  \def\cfstrb{\number\cfllyi.\number\cfllyf}%
  \def\cfstrc{\number\cfurxi.\number\cfurxf}%
  \def\cfstrd{\number\cfuryi.\number\cfuryf}%
  \hbox{\epsfbox[{\cfstra} {\cfstrb} {\cfstrc} {\cfstrd}]{#1.ps}}}
\newcommand {\ehclipfig}[7]{%
  \ifx\cfunit\empty\epsfxsize=#2cm\else\epsfxsize=#2mm\fi%
  \epsfclipon\epsfverbosetrue%
  \cfcmtopspts{#3}\cfllxi=\cftempi\cfllxf=\cftempf%
  \cfcmtopspts{#4}\cfllyi=\cftempi\cfllyf=\cftempf%
  \cfcmtopspts{#5}\cfurxi=\cftempi\cfurxf=\cftempf%
  \cfcmtopspts{#6}\cfuryi=\cftempi\cfuryf=\cftempf%
  \def\cfstra{\number\cfllxi.\number\cfllxf}%
  \def\cfstrb{\number\cfllyi.\number\cfllyf}%
  \def\cfstrc{\number\cfurxi.\number\cfurxf}%
  \def\cfstrd{\number\cfuryi.\number\cfuryf}%
  \ifx\cfunit\empty\hspace{#7cm}\else\hspace{#7mm}\fi%
  \hbox{\epsfbox[{\cfstra} {\cfstrb} {\cfstrc} {\cfstrd}]{#1.ps}}%
  \vspace{-1mm}}
%
%
%
\newdimen\cfllxi \newdimen\cfllyi  \newdimen\cfurxi  \newdimen\cfuryi
\newdimen\cfllxf \newdimen\cfllyf  \newdimen\cfurxf  \newdimen\cfuryf
\newdimen\cftemp \newdimen\cftempi \newdimen\cftempf
\newdimen\cfpspoint \cfpspoint=1bp
%
%
%
\newcommand{\cfcmtopspts}[1]{\ifx\cfunit\empty%
  \cftemp=#1cm\else\cftemp=#1mm\fi%
  \multiply\cftemp10\divide\cftemp\cfpspoint%
  \cftempf=\cftemp\divide\cftemp10\cftempi=\cftemp\multiply\cftemp10%
  \advance\cftempf-\cftemp}
%
%
\def\cfmode{}\def\cfunit{}\clipfigsettings
%

\useunitmm

\def \charthoffset {\hspace{0.2cm}} \def \charthsep {\hspace{0.3cm}}
\def \chartvsepcap {\vspace{0.3cm}}
\def \chartvsep {\vspace{0.1cm}}
\newcommand{\putchartb}[1]{\clipfig{#1}{85}{15}{5}{275}{192}}
\newcommand{\putchartc}[1]{\clipfig{#1}{55}{33}{19}{275}{195}}
\newcommand{\chartlineb}[2]{\parbox[t]{18cm}{\noindent\charthoffset\putchartb{#1}\charthsep\putchartb{#2}\chartvsep}}

\newcommand{\chartlinec}[3]{\parbox[t]{18cm}{\noindent\charthoffset\putchartc{#1}\charthsep\putchartc{#2}\chartvsep\putchartc{#3}\chartvsep}}


\title{A Complete Sample of Soft X-ray Selected AGN: 
II. Statistical Analysis\thanks{Based in part on observations at the European Southern
Observatory La Silla (Chile) with the 2.2m telescope of the Max-Planck-Society
during MPI and ESO time and the ESO 1.52m telescope during ESO time in September
1995 and September 1999.}
}


\author{Dirk Grupe\thanks{Guest observer, McDonald Observatory,
University of Texas at Austin}}
\affil{Astronomy Department, Ohio State University,
    140 W. 18th Ave., Columbus, OH-43210, U.S.A.}
\email{dgrupe@astronomy.ohio-state.edu}




\begin{abstract}
Direct correlations and a Principal Component Analysis (PCA) are
presented for a complete sample of 110 soft X-ray selected AGN of which about half
 are Narrow-Line Seyfert 1 galaxies (NLS1s).
The direct correlation analyses show that narrower FWHM(H$\beta$) correlates with
steeper X-ray spectrum, 
stronger optical FeII emission, weaker [OIII] emission and stronger short-term X-ray
variability. This direct correlation analysis and the PCA 
confirm the
Boroson \& Green (1992) Eigenvector 1 relationship for AGN: FeII strength
anti-correlates with [OIII] line strength.   
Eigenvector 1 is well-correlated with the Eddington luminosity ratio 
$L/L_{\rm Edd}$ while
Eigenvector 2 shows a very good correlation with the mass of the central black
hole \mbh~ and the mass accretion rate \dM.
The Eddington ratio $L/L_{\rm Edd}$ correlates with the X-ray spectral index
\ax~and the black home mass $M_{\rm BH}$ anti-correlates with the X-ray
variability $\chi^2/\nu$.
The Eddington ration $L/L_{\rm Edd}$ may be interpreted as the age of an AGN:
AGN with steep X-ray spectra, strong FeII, and weak [OIII] are AGN in
 an early phase of their evolution.
In this hypothesis NLS1s are young AGN.
\end{abstract}


\section{Introduction}

The ROSAT All-Sky Survey (RASS, \citet{vog99}) led to the discovery of a large
number of previously unknown AGN. While being non-prominent at other wavelengths
these sources are bright at soft X-ray energies. Following the selection
criteria described in \citet{tho98} and \citet{ gru01, gru03a} 
 led to a total of 113 bright soft X-ray
selected AGN. Half of
these AGN are Narrow-Line Seyfert 1 galaxies (NLS1s, \citet{ost85}) which are
characterized by their relatively narrow H$\beta$ emission lines from the Broad
Line Region (BLR), their strong optical FeII emission and weak emission from
forbidden lines from the Narrow Line Region (NLR). NLS1s are the class of
AGN with the steepest X-ray spectra (e.g. \citet{puc92, bol96, bra97,
gru98, gru01, wil03}). 
NLS1 appear to be more common among soft
X-ray selected AGN compared with optically selected samples (e.g. \citet{ste89,
edel99, gru96, gru00a} and references therein).

It has been shown by e.g. 
\citet{lao94, lao97}, \citet{gru96, gru99}, \citet{sul00}, and \citet{vau01} 
 that the width of the BLR
H$\beta$ line, the strength of FeII and the strength of the NLR [OIII] emission
are linked to the X-ray spectral index \ax\footnote{Throughout the paper the 
spectral indeces are  energy spectral induces defined by
$F_{\nu} \propto \nu^{-\alpha}$.}: 
A larger \ax~ goes with an increase
of the FeII strength and decrease of the H$\beta$ width and [OIII] strength.
All these relationships can be described by one basic underlying parameter, often
called Eigenvector 1 or the First Principal Component. 
\citet{bor92} examined the properties of a sample 87 bright PG quasars and found
from a Principal Component Analysis (PCA)
that Eigenvector 1 correlates with the
strength of FeII and anti-correlates with the strength of the [OIII] emission. 
\citet{bor02} and \citet{sul00}
suggested that Eigenvector 1 is the luminosity to Eddington
luminosity ratio $L/L_{\rm Edd}$. This relationship has been clearly
demonstrated by the study of \citet{yuan03} of high-redshift quasars.
\citet{gru96} and \citet{gru99} 
suggested that Eigenvector 1 might represent the 'age' of
AGN and that in this picture NLS1s are AGN in an early phase of their evolution,
a conclusion that has also been drawn for high-redshift quasars and some Broad Absorption
Line Quasars (\citet{mat00, bec00}). The second larges Eigenvector in the
\citet{bor92} sample seems to be the mass accretion rate \dM~ \citep{bor02}.

I have studied a sample of 113 soft X-ray AGN selected from the RASS.
The RASS and pointed ROSAT Position Sensitive Proportional Counter
(PSPC, \citet{pfe87}) and High Resolution Imager
(HRI) observations of this sample
have been described in
detail by \citet{gru01}. 
The objects were selected to be X-ray bright (CR$\geq$0.5 PSPC cts s$^{-1}$) at
high galactic latitude (|b|$>20^{\circ}$)
and
having a soft X-ray spectrum (hardness ratio HR$\leq$0.0; \citet{gru01} and
\citet{gru03a}). The sample is complete for all AGN following these selection
criteria. It is also complete with respect to that X-ray and optical spectra exist
for each source. Even though the sample is X-ray selected it does not miss any
optical counterpart. X-ray and optical properties can be
measured in the same way for each object. 
The sample presented here excludes the known
X-ray transient
sources IC 3599, WPVS007 and RX J1624.9+7554 (\citet{bra95, gru95a, gru95b,
gru99a}) leaving 110 objects in the current sample.
Even though the sample is complete following the selection criteria given 
above, it is biased towards unabsorbed, bright, and
low-redshift sources.  Only a few sources of the sample show
significant optical reddening and X-ray absorption \citep{gru98b}.
The sample misses faint and/or
absorbed X-ray sources. X-ray
absorption by neutral elements not only reduces the observed flux, but because
the sample was selected by hardness ratio, it also converts a source into a hard
X-ray source even though its intrinsic X-ray spectrum may be steep. As shown
recently by \citet{wil03} there is a large number of NLS1s derived from the
Sloan Digital Sky Survey (SDSS, \citet{york00}) which do not show significant
X-ray emission. These sources may be either intrinsically X-ray weak or are
strongly absorbed by neutral elements.  Another bias
comes from the absorption of neutral elements in our Galaxy. Even though the
sample was selected by high-galactic latitude objects, it still misses
borderline objects which may be intrinsically not absorbed, but become 'hard'
X-ray sources
because of the Galactic absorption in the soft X-ray band. 
By extending the sample towards fainter and absorbed sources,
the immediate effect on the sample would be a reduced fraction of NLS1s as
discussed in \citet{gru00a}. However, missing fainter and/or absorbed objects
does not affect the results presented and discussed in this paper.

Optical spectra were obtained at McDonald Observatory,
ESO La Silla, CTIO, and the TLS Tautenburg and the observations, the data
reduction, and the line measurements 
are described in detail in \citet{gru03a} (Paper I). In Paper I we
presented all spectra that have not been published before, performed a simple
statistical analysis of the data and showed the distributions of optical and
X-ray properties among NLS1s and Broad Line Seyfert 1s (BLS1s). NLS1s and BLS1s were
defined following the definition of \citet{good89} with NLS1s being AGN with
FWHM(H$\beta$)$\leq$2000 \kms~and BLS1s with FWHM(H$\beta$)$>$2000 \kms. 

In the soft X-ray selected sample
we found that NLS1s and BLS1s have different distributions 
of their rest frame equivalent width EW(FeII),  
FeII/H$\beta$ ratio, \ax, and X-ray variability. NLS1s have on
average the highest EW(FeII) and FeII/H$\beta$ ratios, the steepest X-ray spectra and
are more variable than BLS1s.  
In this sample they have similar distributions in their
redshifts, luminosities, and EW(H$\beta$). \ax~was
determined from a powerlaw model with Galactic absorption of neutral elements
to the RASS spectra \citep{gru01} with column densities $N_{\rm H, gal}$ 
given by \citet{dic90}. \ax~ is independent of $N_{\rm
H}$ and NLS1s and BLS1s have similar distributions of their Galactic $N_{\rm H}$
values.

The black hole masses were estimated from the monochromatic luminosity at
5100\AA~and FWHM(H$\beta$) listed in Table 2 in \citet{gru03a}
using the relationships given in \citet{kas00} by
determining the size of the BLR by $R_{\rm BLR} \propto~\lambda L_{5100}^{0.7}$ 
and the mass of the
black hole by \mbh~$\propto~R_{\rm BLR}\times FWHM({\rm H}\beta$).   
These
black hole masses were used to determine the Eddington luminosities. The rest frame
bolometric luminosities $L_{\rm bol}$ were estimated from a power law model with
exponential cutoff (see Figure\, 2 in \citet{gru03a}) to the optical and X-ray data.
The Eddington and bolometric luminosities given here are approximate.
An analysis of the black hole masses of the sample
objects shows that for a given luminosity NLS1 have smaller black hole masses
than BLS1s and that the \citet{mag98} and \citet{tre03}
$M_{\rm BH}~-~\sigma$
relationship between the black hole mass and galaxy bulge stellar velocity
dispersion does not apply for most NLS1s (\citet{gru03b}).

In this paper I present the statistical analysis of the complete
sample of 110 soft X-ray selected AGN defined and presented in Paper I.
The paper is organized as follows: In \S\,\ref{analysis} the data analysis,
especially the PCA, is explained, in \S\,\ref{results} the results of the
statistical analysis are presented, followed by a discussion in \S\,\ref{discuss}.
 Luminosities are calculated assuming a Hubble
constant of $H_0$ =75 \kms Mpc$^{-1}$ and a deceleration para\-meter of $q_0$ =
0.0. 

\section{\label{analysis} Data analysis}

The correlation analysis presented in this paper uses the Spearman
rank order and the Student's t-test to check the significance of the
correlation. The advantage of the Spearman rank order analysis compared with a
correlation analysis on the direct measurements
is that outliers do not affect the results as much plus the correlation coefficient 
is not as strongly  
 affected if the correlation is not a linear function.

One main aspect of this paper is the Principal Component Analysis of the
data. The idea behind this method is to reduce the number of parameters
to describe a source from many to a few relevant parameters. The mathematical
formalism behind the PCA is  linear algebra to find the Eigenvalues 
and Eigenvectors for the correlation coefficients matrix.  It is not the
task of this paper to describe this method in detail. An excellent
description of how PCA works can be found in \citet{fra99}.

\section{\label{results} Results}

\subsection{\label{corr_analysis} Correlation Analysis}

Table\,\ref{correlation_tab} contains the Spearman rank order correlation
coefficients between FWHM(H$\beta$) and [OIII]$\lambda$5007\AA, the equivalent
widths (EW)
of H$\beta$, [OIII] and the FeII$\lambda$4570\AA~blend, the flux ratios
[OIII]/H$\beta$ and FeII/H$\beta$,  \ax,  the 0.2-2.0
kev rest-frame X-ray luminosity and the soft X-ray variability parameter 
$\chi^2/\nu$.
There are five rows in the table for each
observational property, in the following order: the whole sample (110 sources),
high-luminosity sources with log $L_{\rm X}>$37.0 [W] (57), low-luminosity 
objects with log $L_{\rm X}<$37.0 [W] (53), NLS1 (51), and BLS1s (59).
The part below the diagonal displays the Spearman rank order correlation
coefficient \rs~
and the part above the diagonal shows the value \ts~ from a Student's
t-test  to indicate the significance of the correlation. 
All correlations with \ts $>$ 3.0 are shown as bold face. 
A \ts=3.0 for 110 source is equivalent to a probability P$<$0.1\% of a random distribution. For
the NLS1s, BLS1s, low-luminosity, and high-luminosity subsamples with their 50-60 sources a
\ts=3.0 is equivalent to P$<$0.2\%.
In the following the most
important correlations are presented.

\subsubsection{\label{fwhm_hb_ax} FWHM(H$\beta$) and \ax}

\begin{figure}
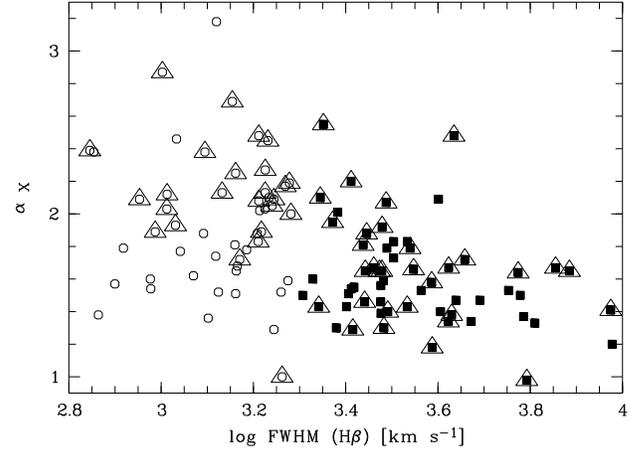

\clipfig{DGrupe.fig1}{87}{15}{5}{275}{192}
\caption{\label{fwhb_ax} log FWHM(H$\beta$) vs. X-ray spectral slope \ax~; 
NLS1 are coded as open circles  and BLS1s and filled squares. 
In addition, high-luminous AGN (log
$L_{\rm X}>$37.0 [W]) are displayed as triangles that surround the circles and
squares.
}
\end{figure}

Figure\,\ref{fwhb_ax} displays the well-known relationship between FWHM(H$\beta$) and
 \ax, namely that objects with steep X-ray spectra
appear to have relatively narrow BLR emission lines.
This relationship has
been established for many optically and X-ray selected samples (e.g.
\citet{bol96, gru96, lao94, lao97, gru99, vau01}). In our 
complete sample the 
FWHM(H$\beta$) and \ax~are well-correlated with a Spearman rank order
correlation coefficient \rs=--0.49 with a Student's t-test \ts=--5.34. 
As noted by \citet{gru99} and \citet{vau01}
this correlation 
improves when only the high-luminosity sources are considered. In the current
 sample, \rs~increases to --0.63 with \ts=--6.01. 
(Table\,
\ref{correlation_tab}). When only the low-luminosity sources are taken into account,
the correlations significantly decreases to \rs=--0.41 with \ts=--3.18.
It is interesting to note that the
FWHM(H$\beta$)-\ax~anti correlation can be seen only when NLS1 and BLS1s are both
included
 When the two groups are examined separately, the
correlations seems to vanish. This is in agreement with the results of
\citet{xu03} for 155 BLS1s for which there was no
correlation found between FWHM(H$\beta$) and \ax~ for their sample. 
The reason is that the ranges in FWHM(H$\beta$ ) become to small to have a large 
enough 'lever-arm' to detect any significant correlation. 

\subsubsection{\label{o3_fe2} [OIII] and FeII}

\begin{figure*}
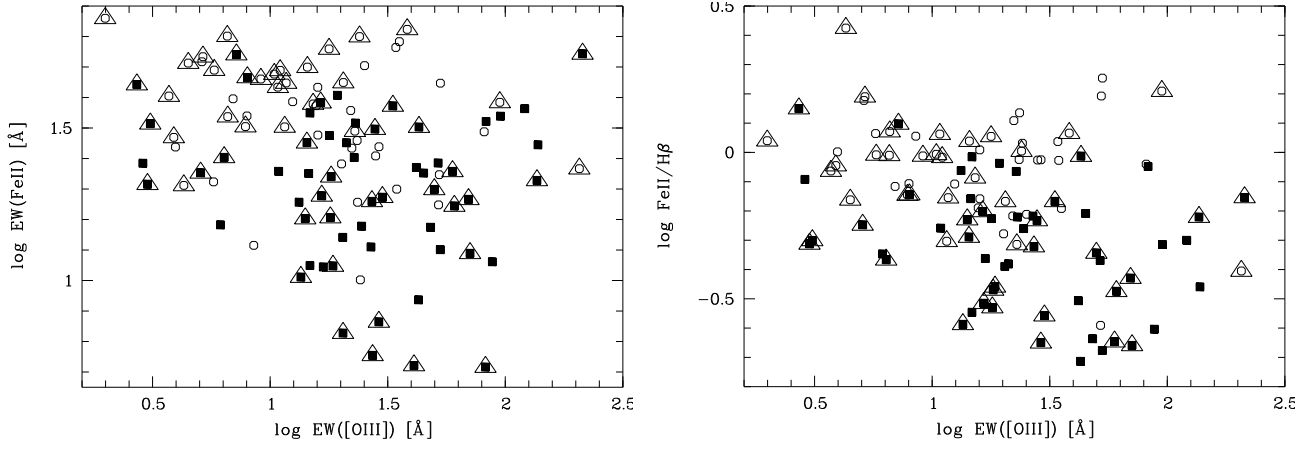

\chartlineb{DGrupe.fig2a}{DGrupe.fig2b}
\caption{\label{o3_fe2_plots} log EW([OIII]) vs. log EW(FeII) (left) and 
log FeII/H$\beta$ (right),  X-ray spectral slope \ax; 
Symbols are  as described in Figure\,\ref{fwhb_ax}.
}
\end{figure*}

\begin{figure}
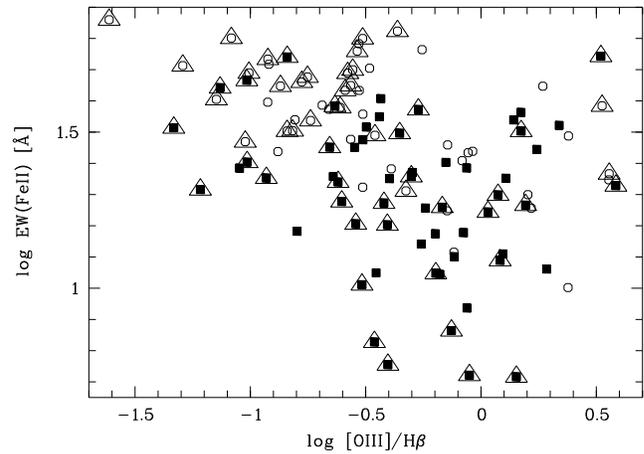

\clipfig{DGrupe.fig3}{87}{15}{5}{275}{192}
\caption{\label{o3hb_ewfe2} log [OIII]/H$\beta$ flux ratio vs. log EW(FeII);
Symbols are as described in Figure\,\ref{fwhb_ax}.
}
\end{figure}

\citet{bor92} found one of the best-known relationships among the observational 
properties of AGN: the strength of the FeII emission anti-correlates with the
[OIII] strength (see also \S\,\ref{pca}). This relation has been found in many optically
selected samples of low-redshift AGN (e.g. \citet{vau01, xu03}) and most recently even in a
sample of high-redshift quasars by \citet{yuan03}. 
In the sample presented here, this trend is clearly confirmed. 
Figure\,\ref{o3_fe2_plots}
shows the anti correlations between the EW([OIII]) and EW(FeII) (left) an the 
FeII/H$\beta$ ratio (right).
For the whole sample the Spearman rank
order correlation coefficients are \rs=--0.28 with \ts=--2.98 and \rs=--0.37 with
\ts=--4.11, respectively. These anti correlations become 
even stronger among the high
luminosity sources (Table\,\ref{correlation_tab}). The [OIII] - FeII anti-correlation is even
stronger between EW(FeII) and the [OIII]/H$\beta$ line flux ratio with \rs=--0.45 and
\ts=--5.23. Figure \ref{o3hb_ewfe2} displays this relation.

\subsubsection{\label{hb_fe2} H$\beta$ and FeII}

\begin{figure*}
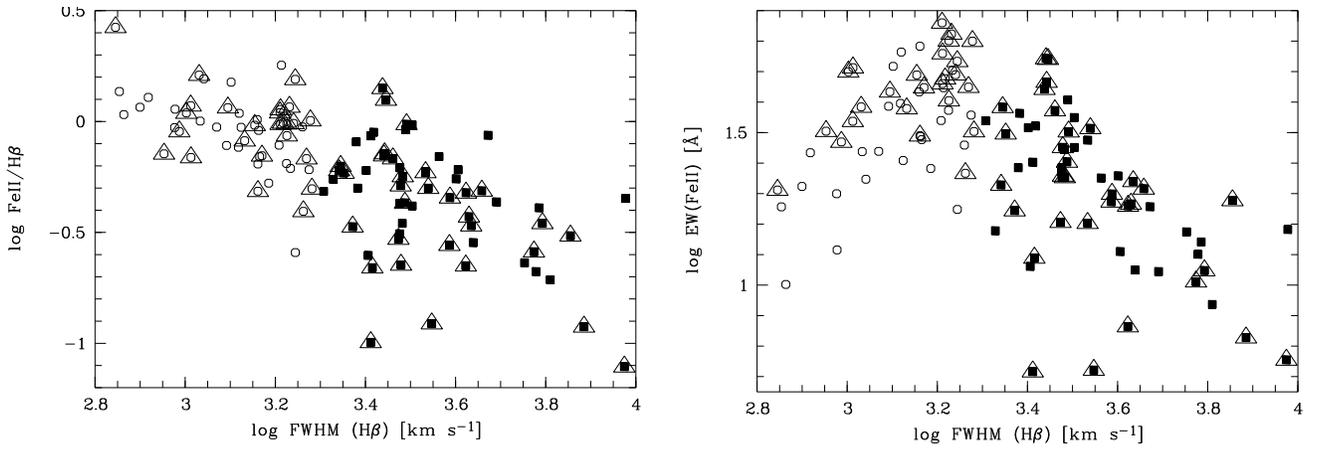

\chartlineb{DGrupe.fig4a}{DGrupe.fig4b}
\caption{\label{fwhm_hb_fe2} log FWHM(H$\beta$) vs. log FeII/H$\beta$ (left) and EW(FeII)
(right);
Symbols are  as described in Figure\,\ref{fwhb_ax}.
}
\end{figure*}

As noticed by e.g. \citet{bor92}, 
\citet{sul02}, and \citet{zheng02} and most recently by \citet{mar03} and
\citet{step03} there
 is an anti-correlation
between the FWHM(H$\beta$) and the EW(FeII) and FeII/H$\beta$  flux ratio. Also the AGN of the
soft X-ray selected sample show these anti-correlations (Figure\,\ref{fwhm_hb_fe2}).
The anti-correlation between FWHM(H$\beta$) and the FeII/H$\beta$ flux ratio is the strongest
among independent
properties in the whole sample with \rs=--0.71 and \ts=--10.5. This anti-correlations
remains this strong even if the sample is split into a low-luminosity and high-luminosity
subsample. The FWHM(H$\beta$) - EW(FeII) anti-correlation is strong for the whole sample with
\rs=--0.49 and \ts=--5.84 and slightly stronger among the high-luminosity objects with \rs=--0.63
and \ts=--6.05.
Interestingly, NLS1s and BLS1s show opposite correlations of
FWHM(H$\beta$) and EW(FeII): while for NLS1s the EW(FeII) increases with H$\beta$ line
widths, the EW(FeII) decreases with H$\beta$ line widths for the BLS1s. For the whole
sample, the right panel in
Figure\,\ref{fwhm_hb_fe2} shows a peak in EW(FeII) at around
FWHM(H$\beta)\approx$2000\kms.

\subsubsection{\ax~ and FeII and [OIII] relations}

\begin{figure*}
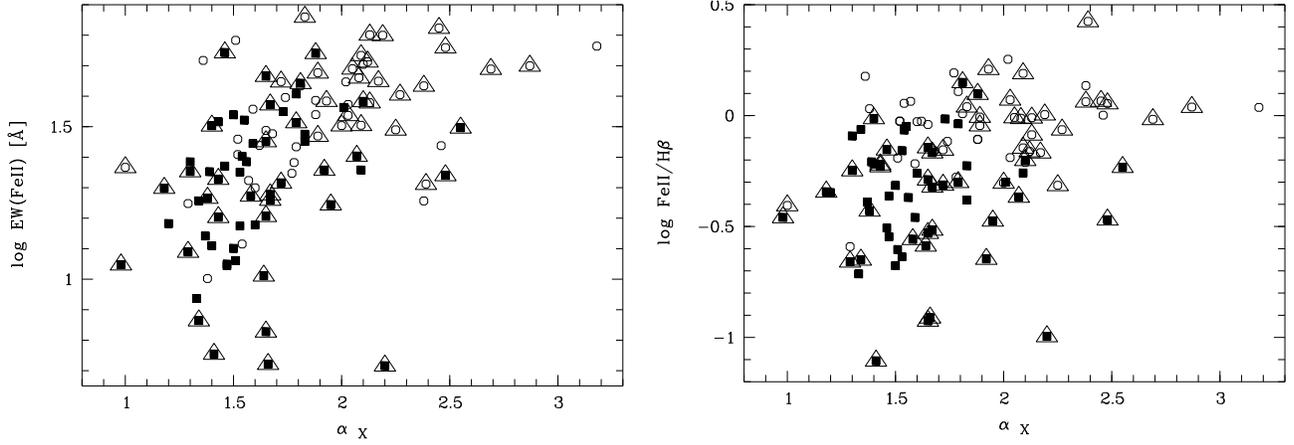

\chartlineb{DGrupe.fig5a}{DGrupe.fig5b}
\caption{\label{ax_fe2_plot} X-ray spectral slope \ax~vs. log EW(FeII) (left) and log
FeII/H$\beta$ (right);
Symbols are  as described in Figure\,\ref{fwhb_ax}.
}
\end{figure*}

\ax~does not only show an anti-correlation with 
FWHM(H$\beta$) (\$\,\ref{fwhm_hb_ax}) it also
correlates with the EW(FeII) and the FeII/H$\beta$ line ratio.
This relation was discovered by \citet{wilkes87} and has been found in a number
of samples e.g. by
\citet{lao94, lao97} and \citet{vau01}. Figure\,\ref{ax_fe2_plot} displays these relations.
The \ax~-EW(FeII) correlation is strong with \rs=+0.55 and \ts=+6.87 for the whole sample and
stays no matter how the sample is split (see Table\/,\ref{correlation_tab}). The correlation
between \ax~and the FeII/H$\beta$ flux ratio is weaker with \rs=+0.40 and \ts=+4.57, but still
significant.

\begin{figure*}
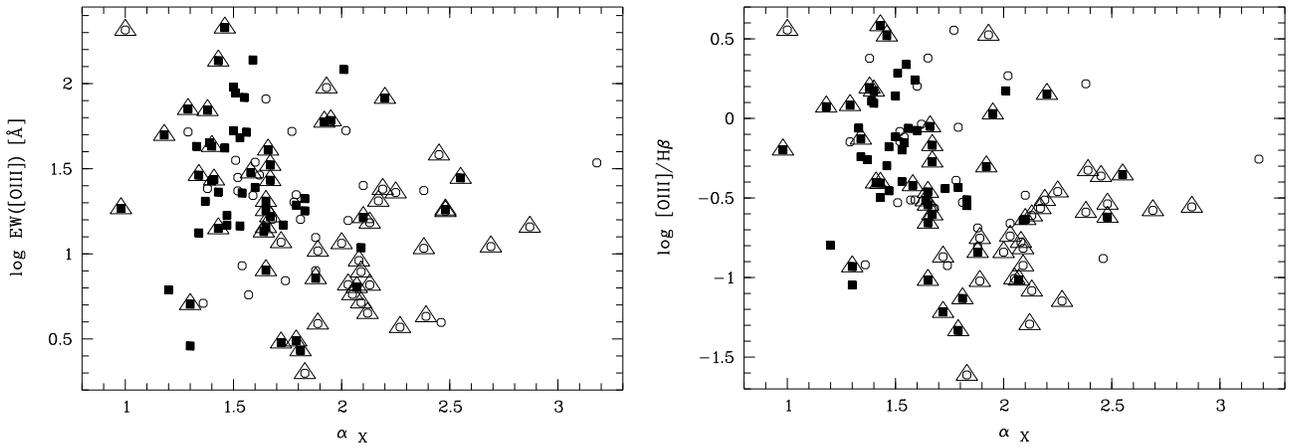

\chartlineb{DGrupe.fig6a}{DGrupe.fig6b}
\caption{\label{ax_o3_plot} X-ray spectral slope \ax~ vs. log EW([OIII]) (left) and log
 [OIII]/H$\beta$ flux ratio (right).  
Symbols are as described in Figure\,\ref{fwhb_ax}.
}
\end{figure*}

As shown is \S\ref{o3_fe2} there is an anti-correlation between the strengths of the
[OIII]$\lambda$5007 line and the 4570\AA FeII blend. This result suggests that \ax~should be
anti-correlated with the [OIII] line strength and indeed 
there is as shown in Figure\,\ref{ax_o3_plot}. 
\ax~ is anti-correlated to the EW([OIII]) with \rs=--0.31 and
\ts=--3.34 and to the [OIII]/H$\beta$ ratio with \rs=--0.40 and \ts=--4.53. A similar result has
been reported previously by e.g. \citet{vau01}.

\subsubsection{\label{loglx_ax_sect} $Log~L_{\rm X}$ vs. \ax}

\begin{figure}
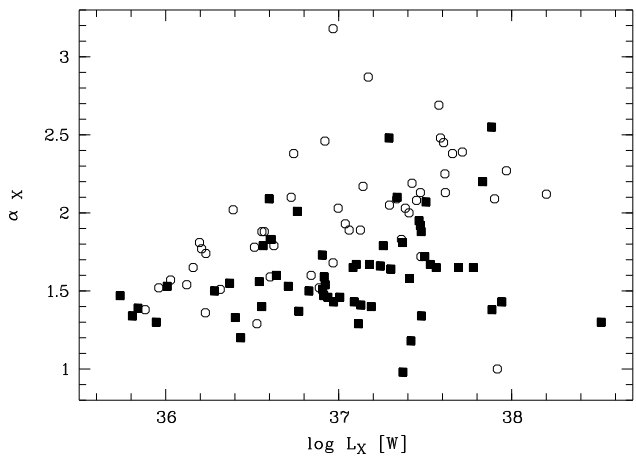

\clipfig{DGrupe.fig7}{87}{15}{5}{275}{192}
\caption{\label{loglx_ax} Soft X-ray rest-frame 0.2-2.0 X-ray luminosity vs.
\ax.
Symbols are as described in Figure\,\ref{fwhb_ax}.
}
\end{figure}

NLS1s show a strong correlation between the rest-frame 0.2-2.0 keV X-ray
luminosity and \ax~(Fig.\,\ref{loglx_ax}) with \rs=0.63 and \ts=5.7. This result
confirms the findings by \citet{lao94} and \citet{gru99} and most recently by
\citet{wil04}. As shown in Fig.\,\ref{loglx_ax} and 
Table\,\ref{correlation_tab}, 
this correlation is not present among BLS1s.

\subsubsection{\label{log_lx_o3} log $L_{\rm X}$ and [OIII]}

\begin{figure*}
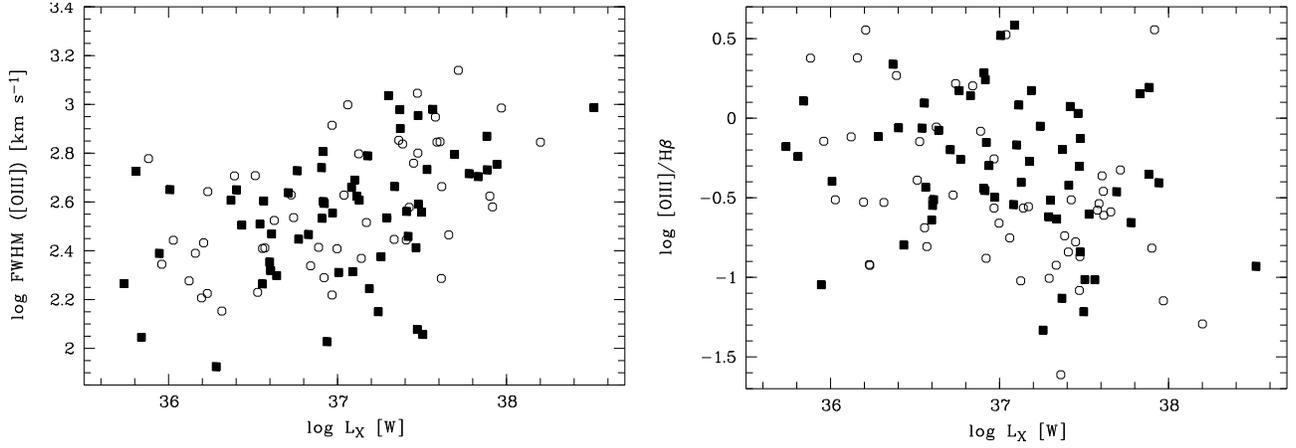

\chartlineb{DGrupe.fig8a}{DGrupe.fig8b}
\caption{\label{loglx_o3} Soft X-ray luminosity log $L_{\rm X}$ 
vs. log FWHM([OIII]) (left) and log
 [OIII]/H$\beta$ flux ratio (right).  
Symbols are as described in Figure\,\ref{fwhb_ax}.
}
\end{figure*}

As noticed by \citet {gru96} and \citet{gru99}, the luminosity of the sources correlates with
the FWHM([OIII]) and anti-correlates with the [OIII]/H$\beta$ line ratio. As displayed in
Figure\,\ref{loglx_o3} these relations appear to be present in the current sample. The
correlation between the 0.2-2.0 soft X-ray luminosity and FWHM([OIII]) is rather strong with
\rs=+0.46 and \ts=+5.41 and appears to be of similar significance among NLS1s and BLS1s (see
Table\,\ref{correlation_tab}). The log $L_{\rm X}$ - [OIII]/H$\beta$ anti-correlation is weaker,
but present with \rs=--0.33 and \ts=--3.60.

\subsubsection{\label{lum_sect} Correlations among luminosities}

\begin{figure*}
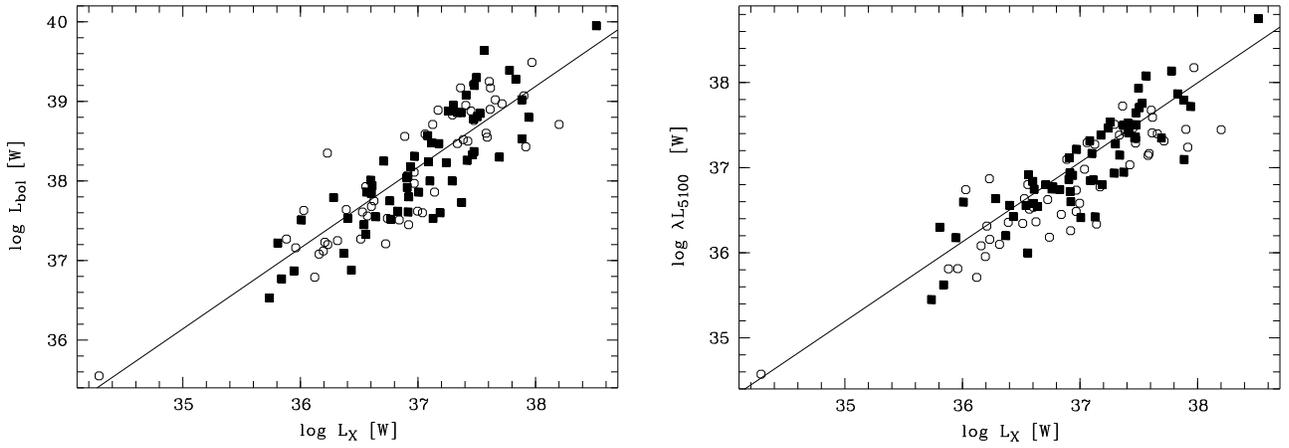

\chartlineb{DGrupe.fig9a}{DGrupe.fig9b}
\caption{\label{lx_lbol_lx_l5100} Rest-frame 0.2-2.0 X-ray luminosity $L_{\rm
X}$ vs. bolometric luminosity $L_{\rm bol}$ (left) and monochromatic luminosity
at (rest-frame) 5100\AA~ $\lambda L_{\rm 5100\AA}$ (right). The solid lines are
the regression curves given in Eqs. 1 and 2 in \S\,\ref{lum_sect}.
Symbols are as described in Figure\,\ref{fwhb_ax}.
}
\end{figure*}

Figure\,\ref{lx_lbol_lx_l5100}
displays the rest frame 0.2-2.0 keV X-ray luminosity
$L_{\rm X}$ vs. the bolometric luminosity $L_{\rm bol}$  and the
rest frame monochromatic luminosity at 5100 \AA~$\lambda L_{\rm 5100\AA}$. 
Figure\,
\ref{lx_lbol_lx_l5100} also shows the regression curves through the data with
the following relationships:

\begin{equation}
log~ L_{\rm bol}~=~(0.44\pm2.30)~+~(1.02\pm0.06) \times log~ L_{\rm X}
\end{equation} 

and

\begin{equation}
log~\lambda L_{\rm 5100\AA}~=~(2.66\pm0.93)~+~(0.93\pm0.05) \times log~L_{\rm X}
\end{equation}

Note that because the EUV part of the AGN Spectral Energy Distribution is
unobservable it leaves large uncertainties in estimating the bolometric
luminosities (e.g. \citet{elv94}). Therefore the equations given above can only
be used for rough estimates.

\subsubsection{\label{chisqu_corr} Correlations with X-ray variability}

\begin{figure*}
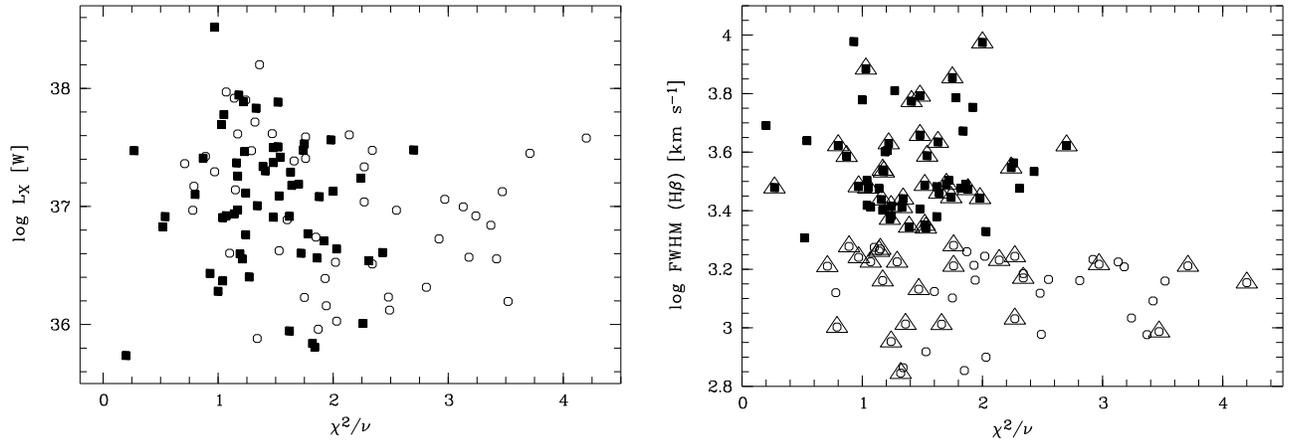

\chartlineb{DGrupe.fig10a}{DGrupe.fig10b}
\caption{\label{chisqu} Soft X-ray variability parameter $\chi^2/\nu$ vs. 
rest-frame 0.2-2.0 X-ray luminosity $L_{\rm X}$ and FWHM(H$\beta$).
NGC 4051, Mkn 766, and RX J1304.2+0205 are of the plots ($\chi^2/\nu$=18.5, 8.86, and
5.08, respectively).
Symbols are as described in Figure\,\ref{fwhb_ax}.
}
\end{figure*}

Figure\,\ref{chisqu}a displays the X-ray variability parameter $\chi^2/\nu$ 
vs. X-ray rest frame X-ray luminosity (see also \citet{gru01}).
For the whole sample there is a light anti correlation with \rs=--0.28
with \ts=--2.81. This result confirms the findings of e.g. \citet{barr86, law93,
green93} and \citet{lei99a} that sources with higher luminosity tend to be less
variable than low-luminosity sources.

Figure\,\ref{chisqu}b shows the anti correlation between 
$\chi^2/\nu$ and FWHM(H$\beta$) for
the whole sample this relationship is anti correlated with \rs=--0.34 with \ts=--3.73. As
shown in the distribution of $\chi^2/\nu$ (Figure\,10 in \citet{gru03a}) NLS1s tend to
be more variable in soft X-rays than BLS1s. This confirms the findings of
\citet{lei99a} that for a given luminosity NLS1s are more variable than BLS1s, and
\citet{wan98} who found a correlation of the FWHM(H$\beta$) with the X-ray 
doubling time scales.

The X-ray variability $\chi^2/\nu$ was derived from the RASS light curves
\citep{gru01}.
Keep in mind that $\chi^2/\nu$ is not only a
function of variability but also of signal-to-noise. Our estimates of the
variability parameter $\chi^2/\nu$ take the errors from the instrumental
background and photon noise into account \citep{gru01}. 
The contribution of noise to the light curve of a
single source depends on how bright the source was and how long the RASS scan
was. Brighter sources observed during a longer RASS scan ($\approx$30s) have an
error from photon noise typically in the order of 10\% while for fainter sources
during a shorter scan ($\approx$10s) the error is typically in the order of
30-40\%.
However, because the sample is selected by count rate and most
sources have count rates between 0.5-2.0 PSPC cts s$^{-1}$ and they were all observed in the
same way during the RASS, the contribution of noise to the source signal is about the same in
each source. This means that noise can affect the result for a single source, 
but not 
statistically for the whole sample. If the variability is due to noise it would
be visible if the sources with lower count rates are more variable (larger
$\chi^2/\nu$) than the sources with larger count rates. 
A check between the RASS
count rate and $\chi^2/\nu$ shows that this is not the case.

\subsection{\label{pca} Principal Component Analysis}

A PCA was performed for the following input parameters: FWHM(H$\beta$), 
FWHM([OIII]), the equivalent widths of H$\beta$, [OIII], and FeII, [OIII]/H$\beta$
and FeII/H$\beta$ flux ratios, \ax, rest-frame 0.2-2.0 X-ray
luminosity $L_{\rm X}$, and the X-ray variability parameter $\chi^2/\nu$.
The results of the PCA are summarized in Table\,
\ref{pca_tab}. The first three Principal Components or
Eigenvectors (EVs) account for more than 2/3 of the intrinsic variation in the data, 
or in other
words, with the first three Eigenvectors, the data are already well-described.
The most important Principal Component,
Eigenvector 1, accounts for 32\% of the variance. It
clearly shows  anti-correlations between the FWHM(H$\beta$) and FWHM([OIII])
and, as in the \citet{bor92} Eigenvector 1 relationship, between [OIII]/H$\beta$ and
FeII/H$\beta$. When Eigenvector 1 increases, the X-ray spectra become steeper,
the FeII/H$\beta$ ratio increases and the [OIII] lines broadens while 
[OIII]/H$\beta$ and FWHM(H$\beta$) decrease.

\begin{figure*}
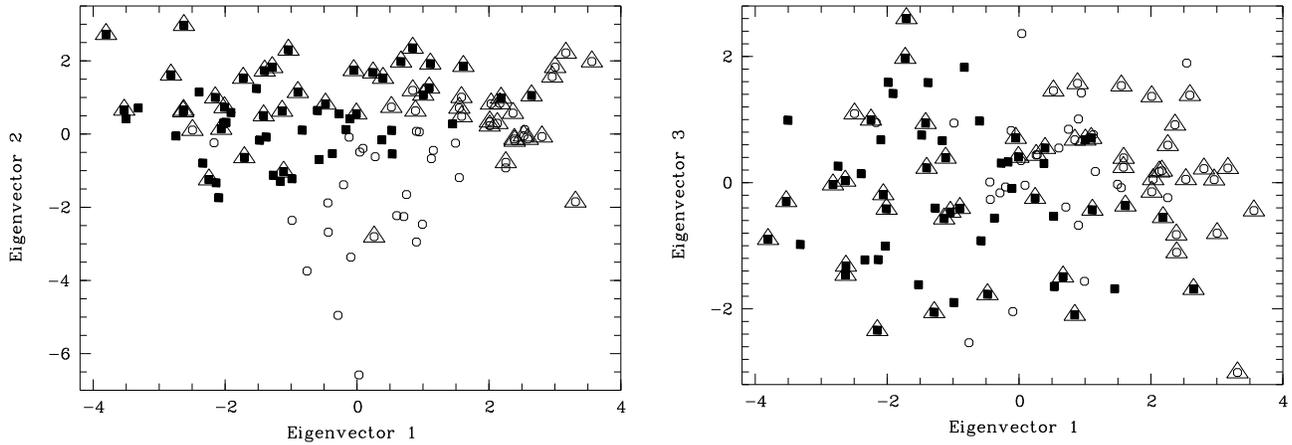

\chartlineb{DGrupe.fig11a}{DGrupe.fig11b}
\caption{\label{pc1_pc2_plot} Eigenvector 1 vs. Eigenvector 2 (left) and
Eigenvector 3 (right)
of the Soft X-ray
selected AGN sample. Symbols are as described in Figure\,\ref{fwhb_ax}.
}
\end{figure*}

Figure\,\ref{pc1_pc2_plot} displays Eigenvector 1 vs. Eigenvector 2  and 
Eigenvector 3. The EV-1 vs. EV-2 diagram (left panel) shows that
NLS1s and BLS1s lie basically
in two different regions, with NLS1s having larger EV-1 values and
lower EV-2 values than the BLS1s. There are a few BLS1s that fall in the NLS1s
regime and vice-versa.

\begin{figure*}
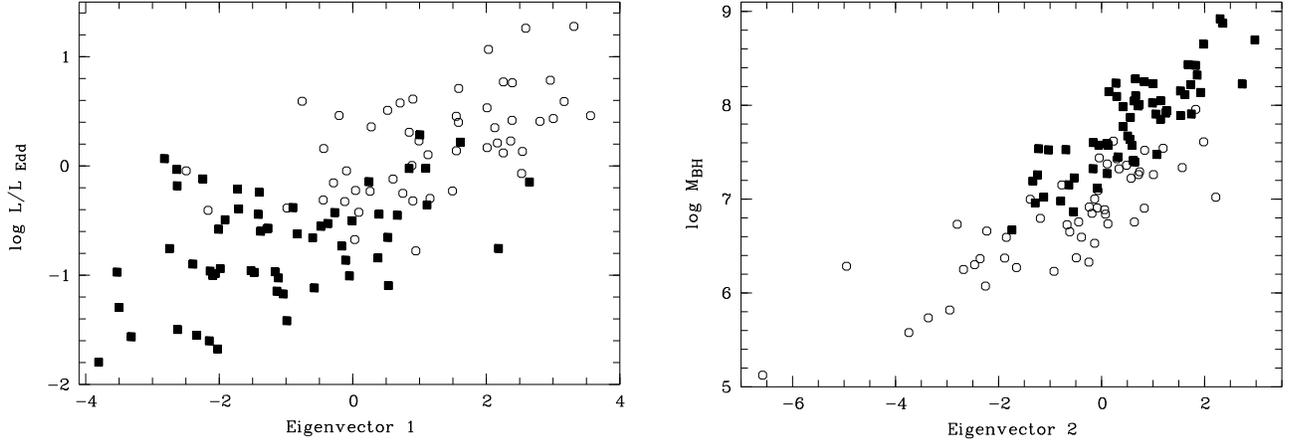

\chartlineb{DGrupe.fig12a}{DGrupe.fig12b}
\caption{\label{pc1_edd_pc2_mbh} Eigenvector 1 vs. Eddington ratio $L/L_{\rm
Edd}$  (left) and Eigenvector 2 (right) vs. black hole mass (right) given in units of
\msun.
Symbols are as described in Figure\,\ref{fwhb_ax}.
}
\end{figure*}

Figure\,\ref{pc1_edd_pc2_mbh} shows the correlation between 
Eigenvector 1 and the
Eddington ratio $L/L_{\rm Edd}$ (left panel) 
and   Eigenvector 2 with the black hole
mass $M_{\rm BH}$ (right panel). 
Table\,\ref{pca_corr} summarizes the Spearman rank order correlation coefficients and
Student's t-test \ts~values of the first two Eigenvectors with the black hole mass
\mbh, the rest-frame 0.2-2.0 keV X-ray luminosity $L_{\rm X}$ and the Eddington ratio
$L/L_{\rm Edd}$. Clearly, EV-1 correlates  with $L/L_{\rm Edd}$ and Eigenvector 2
with \mbh~ and $L_{\rm X}$. On the other hand, EV-1 only shows a marginal correlation 
 with \mbh~and EV-2 does not show any correlation with $L/L_{\rm Edd}$.

\begin{figure*}
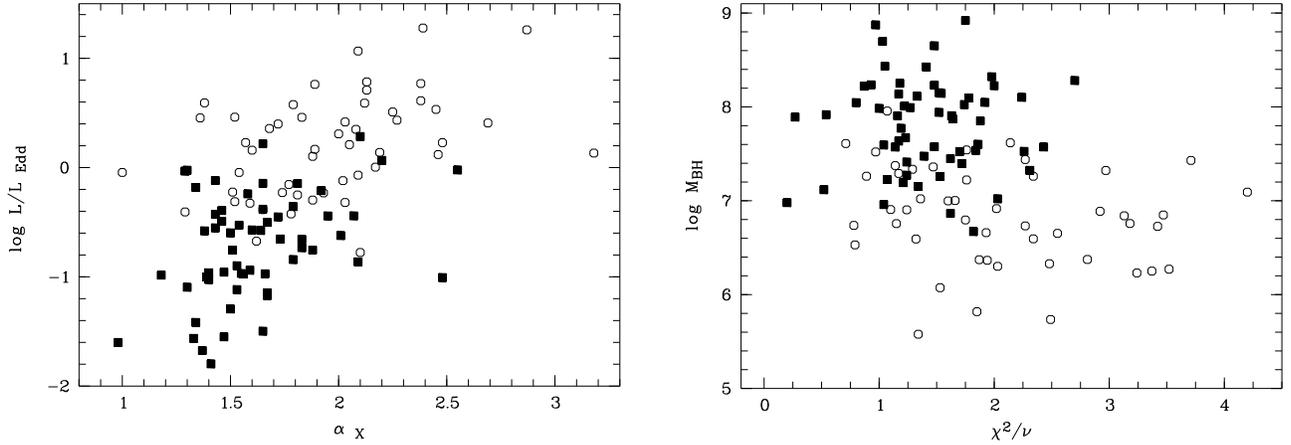

\chartlineb{DGrupe.fig13a}{DGrupe.fig13b}
\caption{\label{ax_edd_chi2_mbh} X-ray spectral index \ax~ vs. Eddington ratio $L/L_{\rm
Edd}$  (left) and  $\chi^2/\nu$ vs. black hole mass (right) given in units of
\msun.
Symbols are as described in Figure\,\ref{fwhb_ax}.
}
\end{figure*}

The correlations of Eigenvector 1 with \ax~and the Eddington ratio
$L/L_{\rm Edd}$ suggest that also \ax~ correlates with $L/L_{\rm Edd}$. As shown in the left
panel of Figure\,\ref{ax_edd_chi2_mbh} this is indeed the case. The correlation is strong with
\rs=+0.54 and \ts=+6.71.

Eigenvector 2 shows an anti-correlation between X-ray luminosity and 
variability $\chi^2/\nu$ and displays the same relationship as shown in
\S\,\ref{chisqu_corr} and Figure\,\ref{chisqu}.
Eigenvector 2 itself is anti-correlated with the X-ray
variability $\chi^2/\nu$ and correlated
with the luminosity and the FWHM(H$\beta$). 
This suggests that also $\chi^2/\nu$  
anti-correlates to the mass of the central black hole as it
has been most recently reported by \citet{papa03}. The right panel in
Figure\,\ref{ax_edd_chi2_mbh} displays this anti-correlation with \rs=--0.39 and \ts=--4.33.
Note that \citet{papa03} used the excess variance (e.g. \citet{nandra97, george00,
lei99a}) to describe the X-ray variability. 

\section{\label{discuss} Discussion}

\subsection{Correlation Analysis}

The sample presented here confirms the 
well-known FWHM(H$\beta$)-\ax~ relationship among AGN
(e.g. \citet{bol96, gru96, lao94, lao97, gru99, vau01}). 
Only NLS1s show very steep X-ray spectra while BLS1s never do. Also this sample
confirms the findings of \citet{gru99} and \citet{vau01} that the
FWHM(H$\beta$)-\ax~relationship is more pronounced among high-luminosity AGN.
Interestingly, this relationship is also true for
 BLS1s samples (e.g. \citet{xu03}). The PCA of the soft X-ray
selected sample shows that EV-1 can be interpreted as the Eddington ratio $L/L_{\rm
Edd}$ and EV-2 possibly by the black hole mass \mbh (see \S\,\ref{pca} and
\S\,\ref{discuss_pca}).
Because a high $L/L_{\rm Edd}$
results in a steep X-ray spectrum (e.g. \citet{pound95}) and a large \mbh~ results in
a
large FWHM(H$\beta$) (e.g. \citet{kas00, pet00}) the FWHM(H$\beta$)-\ax~diagram is
some what similar to the EV-1 EV-2 diagram (Figure\,\ref{pc1_pc2_plot}a). 
This explains
the 'zone of avoidance' in the FWHM(H$\beta$)-\ax~ diagram: sources with broad H$\beta$
lines and therefore large \mbh~would require a mass accretion rate of about 10 \msun
~yr$^{-1}$ to reach their Eddington limit and display a steep X-ray spectrum. Even
though this might be possible for a short time, e.g. during an X-ray outburst (e.g.
\citet{gez03}), it is very unlikely to find a BLS1
in that high outburst state over a long period of time.

Because FWHM(H$\beta)\propto$ \mbh~(e.g. \citet{pet00, kas00}) and
FWHM([OIII])$\propto~ \sigma_*$ (\citet{nel00, shi03}) and the
well-known $M_{\rm BH}~-~\sigma_*$
relationship between the central black hole mass and the bulge stellar velocity
dispersion $\sigma_*$
(e.g. \citet{mag98, geb00, tre03}) one would expect to find a correlation
between FWHM(H$\beta$) and FWHM([OIII]). 
Surprisingly, the Spearman rank order 
coefficient \rs=+0.00
with \ts=+0.01 between FWHM(H$\beta$) and FWHM([OIII]) 
does not confirm this assumption. In a separate paper (\citet{gru03b}) we will
show that  the $M_{\rm BH}~-~\sigma_*$ relationship does not apply for
NLS1s, a result confirming the suggestions of \citet{mat01}, \citet{wan02}, 
and \citet{bian03}.

The strongest (anti-)correlation among independent parameters of the soft X-ray
selected AGN sample is the relation between FWHM(H$\beta$) and FeII/H$\beta$
(\S\,\ref{hb_fe2}). This strong relation suggests that strong FeII emission
occurs in sources with high $L/L_{\rm Edd}$ and strong outflows, as we will
discuss in the following section.

\subsection{\label{discuss_pca} What do the Eigenvectors mean?}
The crucial question of a PCA is whether the Eigenvectors have a physical
interpretation.
At first, one has to keep in mind, that there is not one 'Eigenvector 1' for
every AGN sample. Each individual sample has its own PCA and therefore its own
Eigenvectors.  An
Eigenvector always is specific to a certain sample depending on which observed
parameters have been used and the range of the parameters.
The \citet{bor92}
Eigenvector 1 (meaning the anti-correlation between the FeII and [OIII]
strengths)
does not has to be Eigenvector 1 for all samples. As an example, in
the spectral PCA by \citet{sha03} Eigenvector 1 is the Baldwin effect and the
\citet{bor92} Eigenvector 1 turns out to be Eigenvector 3 in their PG
quasar sample. However, PCAs are comparable to the \citet{bor92} PCA
if the input parameters and the ranges of
the parameters used are similar to those of \citet{bor92}. 

Nevertheless, the EV-1 of the PCA of 
the soft X-ray selected AGN sample is very similar to  the 
\citet{bor92} EV-1.
\citet{bor02} and \citet{sul00}
suggested that for their samples EV-1 can be interpreted as
the Eddington ratio $L/L_{\rm Edd}$. The second Eigenvector 
 EV-2 in the interpretation of \citet{bor02} would be the mass accretion
rate \dM.
In this picture, NLS1 are sources with high Eddington ratio and low mass
accretion rate \dM. The present soft X-ray selected AGN sample seems to confirm these
suggestions. As shown in Table\,\ref{pca_corr}, EV-1 strongly correlates with the
Eddington ratio $L_/L_{\rm Edd}$ and EV-2 with the X-ray luminosity $L_{\rm X}$ and
therefore with the mass accretion rate \dM~($L=\eta * \dot{M} * c^2$). However, 
our EV-2
also strongly correlates with the mass of the black hole. The black hole masses for
the soft X-ray selected sample were estimated by the relationships of \citet{kas00}. 
It has become a popular idea that NLS1s have smaller black hole masses than BLS1s
(e.g. \citet{bol96, wan98}). Also for this soft X-ray selected AGN
sample I can confirm this suggestion (\citet{gru03b}).
As shown in \citet{gru03a}, in our sample the luminosity distributions of NLS1s and
BLS1s are similar.
In order for
an object, like NLS1, with high Eddington ratios to show similar luminosities as
objects with lower Eddington ratios is that their black hole masses are smaller for a
given luminosity than those of BLS1s. However, this does not mean that
individual NLS1s do not have large \mbh. 
As a result of the 
high Eddington ratio, NLS1 are also the sources with the steepest X-ray spectra,
as suggested by e.g. \citet{pound95} and shown is Figure\,\ref{ax_edd_chi2_mbh}.
It is also interesting to note that NLS1s seem to host preferentially in bared
spiral galaxies \citep{cren03} which supposed to have better fueling rates than
normal spirals and would support the result that NLS1s are the sources with the
highest $L/L_{\rm Edd}$. The correlation between \ax~and $L/L_{\rm Edd}$ also
explains the strong dependence of $L_{\rm X}$ on \ax~in NLS1s
(Fig.\,\ref{loglx_ax}). 
The Luminosities of NLS1s are governed by their Eddington ratio while 
in BLS1s their are governed by the mass of the central black hole.  

A high Eddington ratio not only produces steep X-ray spectra
it would also cause an outflow from the central region (e.g.
\citet{laor03} and references therein). \citet{king03} showed recently that when
the outflow rate $\dot M_{\rm out}~\sim~\dot M_{\rm Edd}$ the outflow will be
optically thick and (at least in part) be responsible for the soft X-ray spectra
seen in high accreting objects. Another consequence might be that the outflow
affects the NLR and causes a radial velocity stratification with distance from
the center (e.g. \citet{laor03, smith93}). 
This may explain the increase of the FWHM([OIII])
with increasing Eigenvector 1.

Strong outflows may also be responsible for the strong (anti-)correlation
between FWHM(H$\beta$) and the FeII strength (\S\,\ref{hb_fe2}). Strong optical
FeII emission does not only occur in NLS1s, it also is strong in many Broad
Absorption Line Quasars (BAL QSOs), well-known for their strong outflows (e.g.
\citet{yuan03, hall02, wey91, bor92b}). Because strong FeII emission cannot be
explained by simple photoionization models, \citet{collin00} suggested that the
strong FeII emission found in NLS1s results from shocks which occur in outflows.
This result also explain the correlation between the X-ray spectral index
\ax~and FeII: sources with strong outflows have steep X-ray spectra and strong
FeII emission.

As discussed by \citet{bor92} 
the often-used orientation effect of
Unified Schemes of AGN (e.g. \citet{ant93, urry95}) 
does not explain Eigenvector 1 because
the strength of the [OIII] emission should be independent of orientation if the
NLR is isotropic. This assumption was confirmed for radio-quiet AGN
by the findings of \citet{kur00}
who measured the [OII] and [OIII] emission in a sample of 20 PG quasars and found that
EV-1 correlates with the orientation-independent [OII] and [OIII] emission. As shown
in Table\,\ref{pca_tab} also the PCA of the soft X-ray AGN sample shows correlations
between EV-1 and [OIII]/H$\beta$ and EW([OIII]) suggesting that EV-1 is not an
orientation effect.
One alternative explanation of Eigenvector 1 was mentioned by \citet{bra99} who
suggested that it could be the rotation of the black hole and NLS1 should be
sources with slowly spinning black holes. This was based on the findings of
\citet{bor92} that objects with a large radio loudness have weaker FeII and
stronger [OIII] emission. The driver for radio-loudness is thought to be the
spinning of the black hole (e.g. \citet{wil95}).
However, as \citet{bor92} also remarked, 
a slowly spinning black hole would not explain the properties of ultrasoft
NLS1s. Another argument against this assumption is that meanwhile a number of
radio-loud NLS1 have been found (e.g. \citet{rem91, sie99, gru00b}) which do
not fit into this picture. It has also been shown by e.g. \citet{laor00} and
\citet{lacy01} that there is a correlation between the radio loudness and \mbh.
Because the correlation between EV-1 and \mbh~is found to be marginal
(Table\,\ref{pca_corr}) a link
between EV-1 and the radio loudness is unlikely for the soft X-ray selected AGN
sample.

As shown in Table\,\ref{pca_corr} there are two most likely interpretation of
Eigenvector 2: the black hole mass \mbh~and the mass accretion rate \dM~as suggested by
\citet{bor02}. For both properties, EV-2 shows strong correlations. Arguing for the
black hole mass \mbh~as the interpretation of EV-2 is the correlation with the
FWHM(H$\beta$) (Table\,\ref{pca_tab}). This together with the correlation with 
luminosity and the relationships of \citet{kas00} to determine \mbh~ makes the black hole
mass a plausible explanation of EV-2. However, this does not necessarily explain the
strong weight of the X-ray variability parameter $\chi^2/\nu$ in EV-2. 
On the first view, the X-ray
variability argues more for the mass accretion rate as the interpretation of EV-2.
If the mass accretion rate is low, a small change in the mass accretion rate results
in a higher variability than for objects with high mass accretion rates. 
On the other hand, recently \citet{papa03} reported of a correlation between X-ray variability
and the mass of the central black hole in a sample of 14 AGN supporting the
suggestion that EV-2 represents the black hole mass.
With the
current set of parameters to study the soft X-ray sample it is not possible to give a
final interpretation of EV-2, whether it is \mbh~or \dM.  

The difference between the
PCA of the soft X-ray selected AGN sample and the PG quasar sample of \citet{bor02} is
that his EV-1 is strongly correlated with \mbh~(Table 2 in \citet{bor02}) which is not
the case for the soft X-ray AGN sample (Table\,\ref{pca_corr}). Because in the
\citet{bor02} sample EV-1 as well as EV-2 are correlated with \mbh~the mass accretion
rate \dM~is more likely the explanation for EV-2 than \mbh. However, the slight
correlation of EV-1 with $L_{\rm X}$
(and therefore \dM) in the PCA of the soft X-ray AGN
sample and the weak correlation between EV-1 and \mbh~makes the black hole mass \mbh~
a more plausible interpretation of our EV-2 than \dM.

\subsection{Are NLS1 young AGN?}

The Eddington ratio $L/L_{\rm Edd}$ (and therefore EV-1) can be
interpreted as the 'age' of an AGN. In early phases of the AGN development the
accretion rate would be close to the Eddington limit and decreases at later phases.
 In this picture NLS1 would be AGN in an early phase of the
AGN development (\citet{gru96, gru99, mat00}). One has to be careful with the
term 'age'. Eigenvector 1 or $L/L_{\rm Edd}$ 
do not measure an age in terms of years,
it is more how early in the
evolution of the AGN we see the source, like e.g. T-Tauri stars are stars in an
early state of the stellar evolution while red giants are in a later state. 
A high Eddington ratio in the early phase of the AGN development would not only
produce steep X-ray spectra (e.g. \citet{pound95}) it would also result in strong mass
outflows (e.g. \citet{king03}.
There is growing evidence at optical/UV and X-ray energies
for such strong outflows  in NLS1s
(e.g.  \citet{good00, gru01b, lei97, zam02, sako01, kaspi02, behar03, pounds03}).

\citet{mat00} suggested that NLS1 are low-redshift cousins of high-redshift QSOs and both are
AGN in an early state of their evolution. This assumption recently got
observational support by the findings of \citet{yuan03} from rest-frame optical
spectra of high-redshift QSOs.
They found that a) 
high-redshift QSOs are high accreting sources just like NLS1 and b) they follow the
Boroson \& Green Eigenvector 1 relationship. On the other hand, \citet{con03} argued
on the basis of UV spectroscopy of NLS1 and high-redshift QSOs that they are
not. However, \citet{con03} used UV line parameters and performed a spectral PCA
using the UV data of NLS1 and high-redshift quasars. These are different input
parameters than in the Boroson \& Green Eigenvector 1, like  \citet{yuan03} did
for their sample. 

The reason why about 50\% of the objects in
the soft X-ray AGN sample are NLS1 is simply a selection effect: the objects of the
sample were selected by being X-ray soft and NLS1s are the sources with the
softest X-ray spectra. 
 One crucial point in the interpretation of NLS1
is what their black hole mass really is. For the sample presented here the black
hole masses were derived from the relationship given in \citet{kas00}. This assumes
that the radius of the BLR of NLS1 scales with luminosity in the same way as for
BLS1s. This issue will be discussed in a separate paper
(\citet{gru03b}).

One of the most fundamental relation found in nearby galaxies and AGN is the
black hole mass to bulge stellar velocity dispersion
$M_{\rm BH}~-~\sigma_*$ relation that links properties of the central black hole
with the host galaxy (e.g. \citet{mag98} and \citet{tre03}). However, NLS1s seem
to deviate from this relation as shown by e.g. \citet{mat01, wan02, zheng02} and
\citet{gru03b}. As shown by \citet{gru03b} those objects deviate from the
\citet{tre03} relation that have the highest $L/L_{\rm Edd}$ namely NLS1s. On the
other hand these are the sources with the smallest black hole masses. The
interpretation is that NLS1s are young objects in which the black hole is still
growing and evolving towards the \citet{tre03} $M_{\rm BH}~-\sigma$ relation
\citep{gru03b}.

\subsection{Classification of the Seyfert types}

In Paper I and in this paper we/I adopted the standard definition of 
\citet{good89}
to separate NLS1s and BLS1s. However, even though
this is an easy method, it
fails for a number of sources which have FWHM(H$\beta)>$2000 \kms and are
therefore classified as BLS1s,
but show
all characteristics as NLS1s such as strong FeII emission, weak [OIII] and steep X-ray
spectra and therefore a high EV-1
which puts them into the region of 
NLS1s (Figure\,\ref{pc1_pc2_plot}). Namely these sources are RX J1005.7+4332 and
IRAS 1334+2438. There are also two NLS1s which happen to lie in the BLS1s
region: Mkn 110 and PKS 2227--299, which can also be classified as Seyfert 1.5s
and form a separate subclass.
Future
classification should take the X-ray and optical properties of NLS1s into account to
separate between different Seyfert groups. Even though every separation between these
types are arbitrary because of the continuous properties of the sources, the
classification just on the basis of the FWHM(H$\beta$) is not enough to describe the
classes properly.

\section{Conclusions}

I have presented a statistical analysis of a complete sample of 110 soft X-ray
selected AGN and found:

\begin{itemize}
\item The objects in this sample follow the Boroson \& Green Eigenvector 1 which
can be most likely interpreted as the Eddington ratio $L/L_{\rm Edd}$.
\item Eigenvector 2 of the soft X-ray selected sample is most likely the black hole
mass \mbh.
\item  The Eddington ratio $L/L_{\rm Edd}$ and therefore 
 Eigenvector 1 can be interpreted as  the 
'age' of the AGN.
\item NLS1 in this picture are AGN with a high $L/L_{\rm Edd}$ in a young phase of
their evolution.
\item The soft X-ray AGN sample follows the well-known FWHM$H(\beta$)-\ax~relationship
\item The FeII strength strongly anti-correlates with FWHM(H$\beta$) and
correlates with \ax.
\item The soft X-ray spectral index \ax~correlates with $L/L_{\rm Edd}$ and the X-ray
variability parameter $\chi^2/\nu$ anti-correlates with $M_{\rm BH}$.
\end{itemize}

The sample presented here is complete with respect to the selection criteria
given in the introduction.  Therefore, absorbed, faint and higher-redshift objects are
excluded. Future studies of the sample will be to extend the sample towards
these sources in order to improve the statistical significance of the
correlations.

\acknowledgments
I would like to thank Bev Wills, Karen Leighly,
Stefanie Komossa and Marianne Vestergaard  
for discussions and comments on this paper, and the anonymous referee for
his/her suggestions on the manuscript to improve the final paper.
This research 
has made use of the NASA/IPAC Extra-galactic
Database (NED) which is operated by the Jet Propulsion Laboratory, Caltech,
under contract with the National Aeronautics and Space Administration. 
The ROSAT project is supported by the Bundesministerium f\"ur Bildung
und  Forschung (BMBF/DLR) and the Max-Planck-Society.


\clearpage

\begin{deluxetable}{lcccccccccc}
\tabletypesize{\scriptsize}
\tablecaption{Spearman rank order correlation (below diagonal)
and Student's t-test for the
significance of the correlation (above diagonal). 
 \label{correlation_tab}}
\tablewidth{0pt}
\tablehead{
& \multicolumn{2}{c}{FWHM} & \colhead{EW} & \colhead{EW} & \colhead{EW} & \\
& \colhead{H$\beta$} & \colhead{[OIII]} & \colhead{H$\beta$} &
\colhead{[OIII]} & \colhead{FeII} & \colhead{\rb{[OIII]/H$\beta$}} & 
\colhead{\rb{FeII/H$\beta$}} & \colhead{\rb{\ax}} & 
\colhead{\rb{$log L_{\rm X}$}} & \colhead{\rb{$\chi^2/\nu$}} \\
}
\startdata
& all & +0.01 & +2.02 & +1.88 & {\bf --5.84} & +0.92 & {\bf --10.5} & {\bf --5.34} & +0.61 & {\bf --3.99} \\
& high \lx & {\it --1.30} & {\it +0.47} & {\it +2.31} & {\bf --6.05} & {\it +1.95}  & {\bf  --7.50} & {\bf --6.01} & {\it --0.79} & {\it --0.97} \\
& low \lx &  +0.67 & +2.40 & +0.63 & --2.46 & --0.38 & {\bf --7.50} & {\bf --3.18} & +0.38 & {\bf --4.30} \\ 
\rb{FWHM(H$\beta$)} &  NLS1s & {\it --0.28} & {\bf +5.72} & {\it +1.10} & {\bf +3.30} & {\it --1.27} & {\bf +3.34} & {\it --0.01} & {\it +1.26}  & {\it --1.27} \\
&  BLS1s &  +0.57 & --1.00 & --2.23 & {\bf --4.88} & --1.55 & {\bf --3.35} &  --2.10 & --0.44  & --0.00  \\ \\

& +0.00 & all & +0.23 & --2.79 & +1.80 & --2.42 & +1.66 & +2.54 & {\bf +5.41} & --1.21 \\
& {\it --0.17} & high \lx  &  {\it --1.32} & {\bf --3.45} & {\it +1.40} & {\it --2.42} & {\it +2.49} & {\it +0.62} & {\it +2.47} & {\it +0.67} \\
& +0.09 & low \lx   & --0.13 & +0.59 & +0.33 & +1.53 & +0.50 & +1.20 & +0.90 & --1.37 \\
\rb{FWHM([OIII])} & {\it --0.04} & NLS1s & {\it +0.57} & {\it --1.80} & {\it --1.63} & {\it --0.35} & {\it +1.72} & {\bf +3.70} & {\bf +4.26} & {\it --1.66} \\
& +0.08 & BLS1s  &  --0.40 & --2.41 & +0.38 & --2.21 & +0.72 & +0.75 & {\bf +3.51}  & --0.24  \\ \\

& +0.19 & +0.02 & all & +0.33 & {\bf 3.65} & {\bf --3.46} & {\bf --4.88} & +2.14 & {\bf +3.30} & --2.19 \\ 
& {\it +0.06} & {\it --0.18} & high \lx  & {\it +0.23} & {\it +1.77} & {\it --2.03} & {\it --1.98} & {\it +0.79} & {\it --0.30} & {\it --1.84} \\
& +0.32 & --0.02 & low \lx  & +1.97 & {\bf +3.35} & --2.25 & {\bf --5.60} & +0.60 & {\bf +3.02} & --0.70 \\
\rb{EW(H$\beta$)} & {\bf +0.63} & {\it +0.08} & NLS1s & {\it --0.47} & {\bf +5.35} & {\bf --3.88} & {\bf --5.84} & {\it +0.88} & {\it +2.87} & {\it --2.01} \\
& --0.13 & --0.05 & BLS1s  & +2.11 & +2.21 & --1.06 & --2.92 & {\bf 3.59} & +1.59 & --0.11 \\
 \\	

 & +0.18 & --0.26 & +0.10 & all & --2.98 & {\bf +19.6} & {\bf --4.11} & {\bf --3.34} & --2.04 & --0.72 \\
& {\it +0.30} & {\bf --0.42} & {\it +0.03} & high \lx   & {\bf --3.07} & {\bf +17.5} & {\bf --3.89} & {\it --2.34} & {\it --1.51} & {\it --0.07} \\
& +0.09 & +0.08 & +0.27 & low \lx  & --0.37 & {\bf +9.61} & --2.09 & --0.48 & +0.98 & --1.29 \\
\rb{EW([OIII])} &  {\it +0.16} & {\it --0.25} & {\it --0.07} & NLS1s & {\it --1.39} & {\bf +11.0} & {\it --0.80} & {\it --1.90} & {\it --2.35} & {\it +0.43}\\
& --0.28 & --0.30  &  +0.27 & BLS1s & --0.81 & {\bf +15.1} & --2.58 & --0.96  & --0.89 & --0.07 \\ \\

& {\bf --0.49} & +0.17 & {\bf +0.33} & --0.28 & all   & {\bf --5.23} & {\bf +8.50} & {\bf +6.87} & +1.6 & +1.21 \\
& {\bf --0.63} & {\it +0.19} & {\it +0.26} & {\bf --0.38} & high \lx  & {\bf --4.03} & {\bf +10.7} & {\bf +4.64} & {\it --0.19} & {\it +0.55} \\
& --0.33 & +0.05 & {\bf +0.42} & --0.05 & low \lx  & --2.61 & +2.81 & {\bf +4.52} & +1.24 & +1.77 \\
\rb{EW(FeII)} & {\bf +0.43} & {\it +0.29} & {\bf +0.61} & {\it --0.19} & NLS1s & {\bf --4.30} & {\it +0.59} & {\bf +3.61} & {\it +2.88} & {\it --1.37}\\
& {\bf --0.54} & +0.05 & +0.28 & --0.11 & BLS1s  & --1.74 & {\bf +8.52} &  {\bf +3.15} & +0.05 & --0.06 \\ \\

& +0.09 & --0.23 & {\bf --0.32} & {\bf +0.88} & {\bf --0.45} & all   & --1.90 & {\bf --4.53} & {\bf --3.60} & --0.24 \\ 
& {\it +0.25} & --0.31 & {\it --0.22} & {\bf +0.92} & {\bf --0.48} & high \lx  & {\it --2.91} & {\it --2.58} & {\it 1.20} & {\it +0.03} \\
& --0.05 & +0.20 & --0.30 & {\bf +0.80} & --0.34 & low \lx  & +0.29 & --0.75 & --0.67 & --1.20 \\
\rb{$\rm [OIII]/H\beta$} & {\it --0.18} & {\it --0.23} & {\bf --0.49} & {\bf +0.84} & {\bf --0.53} & NLS1s & {\it +1.20} & {\it --2.18} & {\bf --3.28} & {\it 0.40} \\
& --0.20 & --0.28 & --0.14 & {\bf +0.89} & --0.22 & BLS1s  & --1.19 & {\bf --2.57} &  --1.91 & +0.15 \\ \\

& {\bf --0.71} & +0.16 & {\bf --0.43} & {\bf --0.37} & {\bf +0.63} & --0.18 & all    & {\bf +4.57} & --1.29 & +2.97 \\
& {\bf --0.71} & +0.32  & {\it --0.26} & {\bf --0.46} & {\bf +0.82} & {\bf --0.37} &  high \lx  & {\bf +4.15} & {\it --0.15} & {\it +1.29} \\
& {\bf --0.72} & +0.07 & {\bf --0.62} &  --0.28 & +0.37 & +0.04 &  low \lx  & +2.79 & --1.82 & +2.50 \\
\rb{FeII/H$\beta$} & {\bf --0.42} & {\it +0.24} & {\bf --0.64} & {\it --0.11} & {\it +0.08} & {\it +0.17} & NLS1s & {\it +1.69} & {\it --0.68}  & {\it +0.07} \\
& {\bf --0.40} & +0.10 &  --0.36  & --0.32 & {\bf +0.75} & --0.16 & BLS1s & +0.47  & --1.42 & +0.38 \\ \\

& {\bf --0.46} & +0.24 & +0.20 & {\bf --0.31} & {\bf +0.55} & {\bf --0.40} & {\bf +0.40} & all  & {\bf +4.22} & +1.17 \\
& {\bf --0.63} & {\it +0.08} & {\it +0.11} & {\it --0.30} & {\bf +0.53} & {\it --0.33} & {\bf +0.49} & high \lx  & {\it +1.51} & {\it +0.49} \\
& {\bf --0.41} & +0.17 & +0.08 & --0.07 & {\bf +0.53} & --0.10 & +0.36 & low \lx  & +2.29  & {\bf +3.20} \\
\rb{\ax} & {\it --0.00} & {\bf +0.47} & {\it +0.13} & {\it --0.26} & {\bf +0.46}   & {\it --0.30} & {\it +0.23} & NLS1s & {\bf +5.73} & {--1.30} \\
& --0.27 & +0.10 & {\bf +0.43} & --0.13 & {\bf +0.39} &  --0.32 & +0.06 & BLS1s   & +1.96  & +0.19 \\ \\

& +0.06 & {\bf +0.46} & {\bf +0.30} & --0.19 & +0.16 & {\bf --0.33} & --0.12 & {\bf +0.38} & all  & --2.81 \\
& {\it --0.11} & +0.32 & {\it --0.04} & {\it --0.20} & {\it --0.03} & {\it --0.14} & {\it --0.02} & {\it +0.20} & high \lx  & {\it --1.31} \\
& +0.05 & +0.13 & {\bf +0.39} & +0.14 & +0.17 & --0.09 & --0.25 & +0.30 & low \lx  & --1.40 \\
\rm{log $L_{\rm X}$} &  {\it +0.18} & {\bf +0.52} & {\it +0.38} & {\it --0.32} & {\it +0.38} & {\it --0.42} & {\it --0.10} & {\bf +0.63} & NLS1s & {\it --2.68} \\
& --0.06 & {\bf +0.42} & +0.21 & --0.12 & +0.01 & --0.25 & --0.19 & +0.25 &  BLS1s & --0.71 \\ \\

& {\bf --0.36} & --0.12 & --0.21 & --0.07 & +0.12 & --0.02 & +0.28 & +0.11 & --0.26 & all \\
& {\it --0.13} & {\it +0.09} & {\it --0.24} & {\it --0.01} & {\it +0.07} & {\it +0.01} & {\it +0.17} & {\it +0.07} & {\it --0.17}  & high \lx \\
$\chi^2/\nu$ & {\bf --0.51}  & --0.19 & --0.10 & --0.17 & +0.24  & --0.18 & +0.33 & {\bf +0.41} & --0.19 & low \lx \\
& {\it --0.18} & {\it --0.23} &  {\it --0.28} & {\it +0.06} & {\it --0.19} & {\it +0.06} & {\it +0.01} & {\it --0.18} & {\it --0.36} &  NLS1s \\
& --0.00 &  --0.03 & --0.01 & --0.01 & --0.01 & +0.02 & +0.05 & +0.03 & --0.09 & BLS1s \\
\enddata

\tablenotetext{1}{Listed are
the whole sample (110
objects), AGN with rest-frame 0.2-2.0 X-ray luminosity
$log L_{\rm X}~>~37.0$ [W] (57), AGN with $log L_{\rm X}~<~37.0$ [W] (53), NLS1s
(51), and BLS1s (59). Correlation with a Student's t-test $>$3.0 
are displayed in bold face.
}

\end{deluxetable}

\begin{deluxetable}{lcccccccccc}
\tabletypesize{\footnotesize}
\tablecaption{\label{pca_tab}  Results of the PCA
}
\tablehead{
& \colhead{EV-1} & \colhead{EV-2} & \colhead{EV-3} & \colhead{EV-4} & 
\colhead{EV-5} & \colhead{EV-6} & \colhead{EV-7} & \colhead{EV-8} & 
\colhead{EV-9} & \colhead{EV-10} \\
}
\startdata
Eigenvalue & 3.1973 & 2.3228 & 1.2140 & 1.1593 & 0.8221 & 0.6378 & 0.3887 & 0.2494 & 0.0056 & 0.0028 \\
Proportion & 0.320  & 0.232  & 0.121  & 0.116  & 0.082  & 0.064  & 0.039  & 0.025  & 0.001  & 0.000 \\
Cumulative & 0.320  & 0.552  & 0.673  & 0.789  & 0.872  & 0.935  & 0.974  & 0.999  & 1.000  & 1.000 \\
& \\ \hline \\
log FWHM(H$\beta$) & $-$0.358 & +0.361 & $-$0.134 & $-$0.172 & $-$0.110 & +0.216 & +0.490 & $-$0.629 & $-$0.007 & $-$0.012 \\
log FWHM([OIII])   & +0.229 & +0.195 & $-$0.369 & +0.422 & $-$0.522 & +0.399 & +0.216 & +0.342 & +0.014 & +0.003 \\
log EW(H$\beta$) & $-$0.013 & +0.453 & +0.563 & $-$0.251 & $-$0.159 & +0.187 & $-$0.059 & +0.267 & $-$0.525 & +0.056 \\
log EW([OIII]) & $-$0.393 & $-$0.149 & +0.461 & +0.383 & $-$0.122 & +0.136 & +0.009 & +0.020 & +0.220 & $-$0.618 \\
log EW(FeII) & +0.435 & $-$0.032 & +0.443 & $-$0.027 & +0.102 & +0.459 & +0.041 & $-$0.192 & +0.512 & +0.303 \\
log $\rm [OIII]/H\beta$ & $-$0.373 & $-$0.343 & +0.164 & +0.457 & $-$0.043 & +0.053 & +0.047 & $-$0.092 & $-$0.280 & +0.643 \\
log FeII/H$\beta$ & +0.425 & $-$0.349 & $-$0.035 & +0.122 & +0.212 & +0.286 & +0.097 & $-$0.319 & $-$0.578 & $-$0.331 \\
$\alpha_{\rm X}$ & +0.365 & +0.059 & +0.299 & +0.209 & $-$0.205 & $-$0.650 & +0.508 & $-$0.088 & $-$0.007 & +0.005 \\
log(L$_{\rm X}$) & +0.139 & +0.469 & $-$0.021 & +0.444 & $-$0.030 & $-$0.142 & $-$0.594 & $-$0.433 & $-$0.028 & $-$0.000 \\
$\chi^2/\nu$ &  +0.055 & $-$0.374 & +0.042 & $-$0.344 & $-$0.758 & $-$0.047 & $-$0.292 & $-$0.272 & $-$0.008 & $-$0.002 \\
\enddata
\end{deluxetable}

\begin{deluxetable}{lcccc}
\tabletypesize{\footnotesize}
\tablecaption{\label{pca_corr}  Spearman rank order correlation coefficients \rs~
and Student's T-test values \ts~
for the first two
Eigenvectors with $M_{\rm BH}$, $L_{\rm X}$ and $L/L_{\rm Edd}$. 
}
\tablehead{
   & \multicolumn{2}{c}{EV-1} & \multicolumn{2}{c}{EV-2} \\
   & \colhead{\rs} & \colhead{\ts} & \colhead{\rs} & \colhead{\ts}   
}
\startdata
$M_{\rm BH}$ & --0.29 & --3.20 & +0.80 & +14.0 \\
$L_{\rm X}$ & +0.26 & +2.78 & +0.69 & +9.83  \\
$L/L_{\rm Edd}$ & +0.67 & +9.38 & --0.07 & --0.71  
\enddata
\end{deluxetable}

\end{document}